\documentclass[journal=jpcbfk,manuscript=article,layout=twocolumn]{achemso}
\setkeys{acs}{usetitle = true, etalmode=truncate, maxauthors=5}
\usepackage[version=3]{mhchem} 
\usepackage{graphicx}
\usepackage{bm}

\newcommand{\onlinecite}[1]{\hspace{-1 ex} \nocite{#1}\citenum{#1}} 
\SectionNumbersOn
\author{Kang Kim}
\email{kk@phys.sc.niigata-u.ac.jp}
\altaffiliation{Current address: Department of Physics, Niigata
University, Niigata 950-2181, Japan}
\affiliation[Institute for Molecular Science]
{Institute for Molecular Science, Okazaki, Aichi 444-8585, Japan
}
\alsoaffiliation[The Graduate University for Advanced
Studies]
{School of Physical Sciences, The Graduate University for Advanced
Studies, Okazaki, Aichi 444-8585, Japan}

\author{Shinji Saito}
\affiliation[Institute for Molecular Science]
{Institute for Molecular Science, Okazaki, Aichi 444-8585, Japan
}
\alsoaffiliation[The Graduate University for Advanced
Studies]
{School of Physical Sciences, The Graduate University for Advanced
Studies, Okazaki, Aichi 444-8585, Japan}

\author{Kunimasa Miyazaki}
\altaffiliation{Current address: Department of Physics, Nagoya
University, Nagoya, Aichi 464-8501, Japan}
\affiliation[University of Tsukuba]
{Institute of Physics, University of Tsukuba, Tsukuba 305-8571, Japan}

\author{Giulio Biroli}
\affiliation[CEA Saclay]
{Institut Physique Th{\'e}orique, CEA Saclay, 91191 Gif Sur Yvette,
France and CNRS URA 2306}

\author{David R. Reichman}
\email{drr2103@columbia.edu}
\affiliation[Columbia University]
{Department of Chemistry, Columbia University, New York,
NY 10027, USA}

\title[Dynamic Length Scales in Glass-Forming Liquids]{Dynamic Length Scales in Glass-Forming Liquids: A Inhomogeneous
Molecular Dynamics Simulation Approach}

\keywords{Glass Transition, Dynamic Heterogeneity, Dynamic
Criticallity, Growing Length Scale}

\begin{document}
\begin{abstract}
In this work we numerically investigate a new method for the
 characterization of growing length scales associated with spatially
 heterogeneous dynamics of glass-forming liquids.
This approach, motivated by the formulation of the inhomogeneous
 mode-coupling theory (IMCT) [Biroli G. \textit{et al.}
 \textit{Phys. Rev. Lett.} \textbf{2006} \textit{97}, 195701],
 utilizes inhomogeneous
 molecular dynamics simulations in which the system is perturbed by a
 spatially modulated external potential. 
We show that the response of the two-point correlation function to the
 external field allows one to probe dynamic correlations. 
We examine the critical properties shown by this function, in particular
 the associated dynamic correlation length, that is found to be
 comparable to the one extracted from standardly-employed four-point
 correlation functions. Our numerical results are in qualitative
 agreement with IMCT predictions but suggest that one has to take into
 account fluctuations not included in this mean-field approach in order
 to reach quantitative agreement.
Advantages of our approach over the more conventional one based on
 four-point correlation functions are discussed.
\end{abstract}

\section{Introduction}
\label{Introduction}

In a glass-forming liquid, with decreasing temperature,
the viscosity and the structural relaxation time 
grows rapidly, and
thus the system exhibits a transition to an amorphous
solid~\cite{Wolynes2012Structural}.
The understanding of the underlying mechanism behind this drastic
slowing down is still an open problem in condensed matter physics
even though various theories, experiments, and computer
simulations have been put forth~\cite{Ediger1996Supercooled,
Debenedetti2001Supercooled, Lubchenko2007Theory, Cavagna2009Supercooled,
Berthier2011Theoretical}.
The huge increase in relaxation times occurs
without any obvious structural change~\cite{Ediger1996Supercooled,
Debenedetti2001Supercooled, Lubchenko2007Theory, Cavagna2009Supercooled,
Berthier2011Theoretical}. There is instead a clear change in the dynamics which becomes
increasingly heterogeneous as shown 
by experimental and computational
studies~\cite{Berthier2011Dynamical,Ediger2000Spatially,Hurley1995Kinetic,Kob1997Dynamical,
Yamamoto1997Kinetic}.  
In particular, correlated motion in space and time increases
upon lowering temperatures, resulting in dramatic effects such as violations 
of the Stokes-Einstein relationship~\cite{Berthier2011Theoretical}. The
increasingly large correlations in the dynamics of supercooled liquids
lead to a rather natural definition of growing dynamical
length scales and susceptibilities. Indeed, if one borrows the standard relationship
between order parameter fluctuations and susceptibilities from the theory
of critical phenomena and generalizes the definition of the average order parameter
to be the time-dependent correlation function, e.g. the intermediate
scattering factor, then one is led naturally to the non-linear, 
four-point susceptibility $\chi_{4}(t)$ as the fluctuation
of the order parameter.
Additionally, the 
associated length scale $\xi_{4}$ is quantified from the wave number
dependence of the four-point correlation function $S_4(q, t)$~\cite{Yamamoto1998Dynamics, Franz2000On,
Donati2002Theory, Lacevic2002Growing, Lacevic2003Spatially,
Berthier2004Time, Whitelam2004Dynamic, Toninelli2005Dynamical,
Chandler2006Lengthscale, Szamel2006Four, Berthier2007Spontaneous,
Berthier2007Spontaneous2, Stein2008Scaling, Karmakar2009Growing,
Karmakar2010Analysis, Flenner2010Dynamic, Flenner2011Analysis,
Mizuno2011Dynamical,
Kim2013Multiple,Berthier2005Direct,DalleFirrier2007Spatial,
Brambilla2009Probing, Bouchaud2005Nonlinear, Tarzia2010Anomalous,
CrausteThibierge2010Evidence, Diezemann2012Nonlinear}.

Given that $\chi_{4}(t)$ measures the spatial correlations in the equilibrium 
relaxation process, it is natural to ask whether one can define a related response
function, again trying to follow the usual route set up in critical phenomena. 
Intuitively, a local static perturbation should affect the dynamics far away if the system 
is indeed dynamically correlated. It was indeed shown that within mode-coupling theory
this is the case: the dynamical response function of a supercooled liquid
to a {\it spatially modulated} external field yields interesting information
pertaining to the length scales of dynamical heterogeneity~\cite{Biroli2006Inhomogeneous}. 
Unlike the case of standard critical phenomena, there is no direct fluctuation-dissipation relation
between the equilibrium fluctuations of the dynamical order parameter, $\chi_4(t)$,
and the susceptibility calculated from the dynamical response of the
system to an external field that couples to the fluid density~\cite{Berthier2007Spontaneous,
Berthier2007Spontaneous2}.
The latter quantity formally is related to a three-point, as opposed to
four-point, correlation function. 
This object, distinct from the standard $\chi_{4}(t)$ of traditional studies,
can be given the physical interpretation as a measure of the ``compressibility''
of dynamical trajectories perturbed by an external field modulated
at a fixed length scale.

Because inhomogeneous molecular dynamics simulations are considerably
more difficult to perform reliably than homogeneous ones, one may ask whether there is 
any benefit in calculating dynamic heterogeneity length scales from the dynamical response
functions alluded to above.
In fact there are several compelling reasons
to undertake the study of such quantities. 
First, it has been argued on rather
general grounds that the correlation functions and susceptibilities evaluated
from the response to an external field are less ambiguous with respect to 
{\em ensemble dependence}~\cite{ Berthier2007Spontaneous,
Berthier2007Spontaneous2}. 
Indeed, the ensemble dependence of quantities
associated with $\chi_{4}(t)$ has be the root cause of some degree of confusion
related to the interpretation of dynamical heterogeneity length
scales~\cite{Stein2008Scaling, Karmakar2009Growing,
Karmakar2010Analysis, Flenner2010Dynamic, Flenner2011Analysis}.  
Specifically, the behavior of $\chi_4(t)$ obtained as the $q\rightarrow 0$
limit of $S_4(q, t)$ is quite intricate because it is 
expected to be given by two terms: one term is the $\chi_4(t)$ determined in the
specific ensemble and
the other term includes the correlations of fluctuations which are suppressed in the
chosen ensemble.
Second, an untested prediction  put forward on rather general grounds is that the length
scales associated with the two distinct formulations discussed above are identical~\cite{Berthier2007Spontaneous,
Berthier2007Spontaneous2}.
This prediction deserves scrutiny.
Third, the formulation of approximate microscopic
theories of dynamical heterogeneity such as the ``inhomogeneous mode-coupling theory''
(IMCT) of Biroli {\em et al.} provides quantitative predictions on the
dynamical response to spatially modulated external
fields~\cite{Biroli2006Inhomogeneous,Szamel2010Diverging}. 
A simulation of such function in model
supercooled liquids would provide the first {\em direct} test of the predictions of IMCT.
A final reason for undertaking a study of dynamical heterogeneity from consideration
of the response to an external field is that---if indeed it turns out that this approach
may reliably yield information associated with dynamical heterogeneity---such an avenue may
be a fruitful means for extracting dynamical length scales in experiments, 
in particular in colloidal systems.  The use of laser tweezer technology would allow for a 
facile route of dynamical heterogeneity length scales in systems for which 
$\chi_{4}(t)$ may be difficult or impossible to estimate~\cite{Curtis2002Dynamic}.

The aim of this work is to establish by numerical simulations that the
dynamical response function introduced in the context of IMCT is indeed
able to probe dynamical correlations. 
We shall analyze its critical properties, in particular the dynamic
correlation length that can be extracted from it, and compare them to
their counterparts obtained by usual four point functions. Finally, we
will also contrast our numerical findings to the prediction of IMCT thus
performing the first test of this theory. 

This paper is organized as follows.  In Sec.~\ref{NEMD}, we discuss how we 
obtain the relevant three-point correlation functions from inhomogeneous molecular dynamics (IMD) simulations. 
In Sec.~\ref{IMCT} we recall the main predictions of IMCT that we will test later.
In Sec.~\ref{Model}, 
we introduce our model of supercooled liquids and the techniques used in our
MD simulations and summarize several time scales characterized from 
conventional intermediate scattering correlation functions that are important for later discussion.  
Section~\ref{Results} describes the numerical behavior of multi-point correlation functions from our 
MD simulations. In Sec.~\ref{Fourpoint}, we first summarize the results of the four-point 
correlation functions calculated from standard equilibrium MD simulations. In 
Sec.~\ref{Threepoint}, we present the numerical results for the three-point
correlation functions calculated from the IMD simulations. In each subsection, the dynamic length scales are 
quantified and their temperature and time-scale dependencies are examined.
Comparisons with the predictions of IMCT are also made. In Sec.~\ref{Summary}, we summarize 
our numerical results regarding the dynamical heterogeneity length scale.

\section{general development}
\label{NEMD}

Following the setup of IMCT,
let us consider an $N$-particle system in the presence
of an external field~\cite{Biroli2006Inhomogeneous}.
The total Hamiltonian is described by
\begin{equation}
H = H_0 + U,
\label{Hamiltonian}
\end{equation}
where $H_0$ is the unperturbed Hamiltonian and $U$ is the external potential.
Here we consider an inhomogeneous external field, which is coupled with the
spontaneous density field,
\begin{equation}
U=h \rho_{\bm{q}}(t)=h \sum_{i=1}^{N}\exp[-i\bm{q}\cdot\bm{r}_i(t)],
\end{equation}
with the wave vector $\bm{q} = 2\pi (n_x, 0, 0) / L$.
Here $L$ is the linear dimension of the system and $n_x$ is an
arbitrary nonzero integer.

To provide the information regarding dynamical correlations, the
quantity of interest is 
the response of the two-point correlation function,
\begin{equation}
F(\bm{k},
\bm{q}, t) = \frac{1}{N}\langle \rho_{\bm{k}}(t)
\rho_{-\bm{k}-\bm{q}}(0)\rangle_U,
\end{equation}
where $\langle \cdots \rangle_U$ denotes the equilibrium ensemble average for the
system subject to the weak external potential $U$.
The deviation of the two-point correlation $F(\bm{k}, \bm{q}, t)$ due
to the inhomogeneous external potential $U$ yields the IMCT susceptibility, which is defined as
\begin{equation}
\chi_U(\bm{k}, \bm{q}, t) = -\frac{\mathrm{d} F(\bm{k},
\bm{q}, t)}{\mathrm{d}h}.
\label{chi_u}
\end{equation}

This can be shown to be related to a three-point correlation function:
\begin{align}
\chi_U(\bm{k}, \bm{q}, t) &=
\frac{1}{N}\langle\rho_{\bm{q}}(t)\rho_{\bm{k}}(t)\rho_{-\bm{k}-\bm{q}}(0)\rangle_{\rm  eq}\nonumber\\
&- \int_0^t \mathrm{d}t' \frac{k_BT}{N}\langle
\rho_{\bm{k}}(t)\{\rho_{\bm{q}}(t'), \rho_{-\bm{k}-\bm{q}}(0)\}
\rangle_{\rm eq}.
\label{linear_response_chi_u}
\end{align}
Note that $\{, \}$ is the Poisson bracket and
$\langle\cdots\rangle_{\rm eq}$ expresses the equilibrium ensemble average
in the unperturbed system.
However, the second term in Eq.~(\ref{linear_response_chi_u})
is numerically
demanding because the Poisson bracket should be evaluated by the time
evolution of the classical stability matrix.

\begin{figure*}[t]
\includegraphics[width=.3\textwidth]{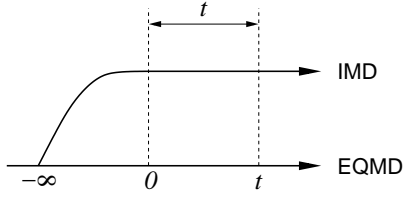}
\caption{
Schematic illustration of the calculation procedure for the
three-point correlation function $\chi_U(\bm{k}, \bm{q}, t)$
via the hybrid IMD and EQMD simulations. The IMD
line represents the ramping profile for the external
force of Eq.~(\ref{eq_external_force}).
}
\label{nemd_schematic}
\end{figure*}

To avoid this hindrance, we explicitly calculate the response
function, $\chi_U = -\mathrm{d}F/\mathrm{d}h$, via the
IMD simulations, in which the external field is applied to the system.
In the course of IMD simulations, the $i$-th particle is subjected to an
external force,
\begin{equation}
\bm{F}_i^{\rm ext} =  -\frac{\partial [2h
  \cos(\bm{q}\cdot\bm{r}_i)]}{\partial \bm{r}_i},
\label{eq_external_force}
\end{equation}
which has been turned on at $t=-\infty$.
Note that the strength of the potential $h$ should be chosen in the
linear response regime.
After reaching steady state at some simulation time,
the numerical data should be recorded for the calculations of
$F(\bm{k}, \bm{q}, t)$.
In practice, Eq.~(\ref{chi_u}) is numerically evaluated using
\begin{equation}
\chi_U(\bm{k}, \bm{q}, t) = - [F(\bm{k}, \bm{q}, t)_{h} - F(\bm{k},
\bm{q}, t)_{h=0}] / h,
\label{eq_chiu}
\end{equation}
where the first term is calculated from the IMD simulations
in which the external field is applied, whereas the second term is
obtained
from the unperturbed equilibrium MD (EQMD) simulations.
This procedure is the well-known subtraction
technique~\cite{Ciccotti1979Thought}.
A schematic of the numerical calculation is illustrated
in Fig.~\ref{nemd_schematic}.
Here we note that the term $F(\bm{q}, \bm{k}, t)_{h=0}$ in
Eq.~(\ref{eq_chiu}) should be exactly zero at any wave numbers $k$
because of the momentum conservation.

\section{IMCT predictions}
\label{IMCT}

In the following we recall the main predictions of IMCT concerning the
function $\chi_U$ when temperature approaches the MCT transition (taking
place at $T=T_{MCT}$).

In the $\beta$ regime one obtains:
\begin{equation}
\chi_U(\bm{k}, \bm{q}, t)=\frac{1}{\sqrt{\varepsilon}+\Gamma q^2}S(k)h(k) \, g_{\beta}\left(q\xi ,
\frac{t}{\tau_{\beta}}\right)
\end{equation}
where $h(k)$ is the critical amplitude, $S(k)$ the structure factor,
$\tau_\beta=\epsilon^{-1/2a}$ (we use the standard MCT notation) and
$g_\beta$ a scaling function. The length-scale $\xi$ diverges as
$\epsilon^{-1/4}$
with $\epsilon=(T-T_{MCT})/T_{MCT}$. The behavior of  $\chi_U(\bm{k},
\bm{q}, t)$ for $q\xi\gg 1$ (but with $q$ still much less than the
wave-vectors corresponding to the microscopic structure) is a power law
as in standard critical phenomena:
\begin{equation}
\chi_U(\bm{k}, \bm{q}, \tau_{\beta})\propto q^{-2}.
\end{equation}
In consequence,
 using the same notation of second order phase transitions, the critical
 exponents for the $\beta$ regime are $\nu=0.25$, $\eta_{\beta}=0$,
 $z_\beta=2/a$~\cite{Biroli2006Inhomogeneous}.

For the $\alpha$ regime IMCT predicts:
\begin{equation}
\chi_U(\bm{k}, \bm{q}, t)=\frac{\Xi(q\xi)}{\sqrt{\epsilon}(\sqrt{\epsilon} + \Gamma q^2)} 
\, g_{\alpha,k}\left(\frac{t}{\tau_\alpha}\right),
\end{equation}
with $\Xi$ a certain regular function with $\Xi(0) \neq 0$ and $\Xi(v
\gg 1) \sim 1/v^2$ such that for $q\xi\gg 1$ (but with $q$ still much
less than the wave-vectors corresponding to the microscopic structure)
\begin{equation}
\chi_U(\bm{k}, \bm{q}, \tau_{\alpha})\propto q^{-4}.
\end{equation}
In consequence,
the critical exponents for the $\alpha$ regime are $\nu=0.25$,
$\eta_{\alpha}=-2$, $z_\alpha=2/a+2/b$.

The matching between the two regimes is given by the small argument
behavior of the function $g_{\alpha,k}(u \ll 1) = S(k)h(k)u^b $. This
implies that as a function of time the growth of $\chi_{{\bf q}}({\bf
k},t)$ scales as $t^b$. The scaling of the correlation length with
$\epsilon$ between $\beta$ and $\alpha$ regime does not change but
the amplitude of $\chi_U$ increases; this suggests that while keeping a
constant spatial extent, the geometrical structure of the dynamic
correlations significantly fatten between $\tau_\beta$ and $\tau_\alpha$
\cite{Biroli2006Inhomogeneous}.

\section{Model}
\label{Model}

\begin{figure*}[t]
\includegraphics[width=.6\textwidth]{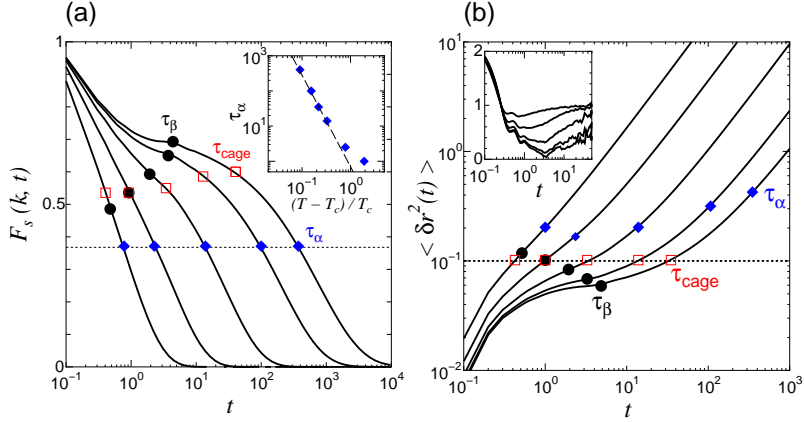}
\caption{
(a) Intermediate scattering function $F_s(k, t)$ with $k=2\pi$ and
(b) Mean square displacement $\langle\delta r^2(t)\rangle$
at temperatures $T=0.772$, $0.473$, $0.352$, $0.306$, and $0.289$ from left to right
for the model system described in Sec.~\ref{Model}.
The time scales $\tau_\beta$, $\tau_{\rm Int}$, and $\tau_\alpha$
are indicated by circles, squares, and diamonds, respectively.
These time scales are described in Sec.~\ref{Model}.
The dotted line represents (a)
$F_s(k, t) = 1/e$ and 
(b) $\langle\delta r^2(t)\rangle = 0.1$, respectively.
Inset of (a): $\alpha$-relaxation time $\tau_\alpha$ as a function of
the temperature $(T-T_c)/T_c$ with $T_c=0.265$.
The dashed line refers to power-law behavior as $\tau_\alpha \sim
 (T-T_c)^{-\gamma}$ with $\gamma = 2.6$.
Inset of (b): $\mathrm{d} \ln
\langle\delta r^2(t) \rangle / \mathrm{d} \ln t$
at temperatures $T=0.772$, $0.473$, $0.352$, $0.306$, and $0.289$ from top to bottom.
}
\label{msd}
\end{figure*}

We have performed MD simulations for a glass-forming
binary soft-sphere mixture~\cite{Bernu1985Molecular, Bernu1987Soft}.
Our system consists of $N_1=10,000$ and $N_2=10,000$ particles of
components 1 and 2, respectively.
They interact via a soft-core potential given as
\begin{equation}
v_{ab}(r)=\epsilon_0\left(\frac{\sigma_{ab}}{r}\right)^{12},
\end{equation}
where
$\sigma_{ab}=(\sigma_a+\sigma_b)/2$ and $a, b \in \{1, 2\}$.
The interaction was truncated at $r = 3\sigma_1$.
The size and mass ratios were taken to be $\sigma_1 / \sigma_2 = 1/1.2$ and $m_1 / m_2
= 1 / 2$, respectively.
The total number density was fixed at
$\rho=(N_1+N_2)/L^3=0.8\sigma_1^{-3}$ with the
system length $L=29.24\sigma_1$ under periodic boundary conditions.
In this study, the numerical results will be presented in terms of reduced
units $\sigma_1$, $\epsilon_0/k_B$, and 
$\tau=\sqrt{m_1\sigma_1^2/\epsilon_0}$ for length,
temperature, and time, respectively.
The velocity Verlet algorithm was used with a time step of $0.005\tau$ in
the microcanonical ensemble.
The investigated thermodynamic states were $T\in[0.772, 0.289]$.
At each temperature, the self-part of the intermediate scattering function for the component
$1$, $F_s(k, t)$, is calculated 
with the wave number $k_{\rm max}=2\pi/\sigma_1$, at which the static
structure factor of component $1$ takes its first peak.
The $\alpha$-relaxation time $\tau_\alpha$ is
determined from the criterion $F_s(k_{\rm max}, \tau_\alpha)=e^{-1}$,
as shown in Fig.~\ref{msd}(a).
The ``mode-coupling'' transition temperature $T_c$ is evaluated from the power law
behavior as $\tau_\alpha\sim(T-T_c)^{-\gamma}$ with $T_c\simeq 0.265$ and
$\gamma\simeq 2.6$.
The relative temperature distance from $T_c$ is given as $\epsilon =
(T-T_c)/T_c \in [1.91, 0.09]$.
To determine the smaller time scale of the $\beta$-relaxation, 
we define the time $\tau_\beta$ at which the function $\mathrm{d} \ln
\langle\delta r^2(t) \rangle / \mathrm{d} \ln t$ has the minimum
value~\cite{Stein2008Scaling}.
Here $\langle\delta r^2(t) \rangle$ is the mean
square displacement for component $1$.
Furthermore, we determine the intermediate time scale, that is referred
to as $\tau_{\rm Int}$, between the two time scales $\tau_\beta$ and $\tau_\alpha$.
This time scale $\tau_{\rm Int}$ is determined from the criterion as
$\langle\delta r^2(\tau_{\rm Int}) \rangle = 0.1{\sigma_1}^2$.
As observed in Fig.~\ref{msd}(b), after this time
the tagged particle can escape from the cage
composed of neighboring particles,
particularly at lower temperatures.
On the other hand, at high temperatures, $\tau_{\rm Int}$ is
approximately equal
to $\tau_{\beta}$.
Figure~\ref{msd}(b) shows the time dependence of the MSD, where
three time scales $\tau_\beta$, $\tau_{\rm Int}$, and $\tau_\alpha$
are shown at each temperature.

In the IMD,
we have performed simulations in the linear response regime with $h=0.02$.
After long time simulations comparable to the
$\alpha$-relaxation time $\tau_\alpha$, the density field
$\rho(\bm{r}, t)$ reaches a stationary state following the profile of $\cos(\bm{q}\cdot\bm{r})$.
Then, the correlation function Eq.~(\ref{eq_chiu}) was calculated.
We averaged the results over 30 independent simulation runs.
The simulation time at lowest temperature is as
long as $t=100,000$.

\section{Results and Discussion}
\label{Results}

\subsection{Four-point correlation function $S_4(q, t)$}
\label{Fourpoint}

We first summarize the numerical results by using
the four-point correlation functions obtained from the EQMD simulations.
We follow previously established work~\cite{Yamamoto1998Dynamics,
Lacevic2002Growing, Berthier2004Time, Berthier2007Spontaneous}, 
with the four-point correlation function $S_4(q, t)$ defined as
\begin{equation}
S_4(q, t) = \frac{1}{N}\langle Q(\bm{q}, t) Q(-\bm{q}, t)\rangle,
\end{equation}
\begin{equation}
Q(\bm{q}, t) =\sum_{i=1}^N W_i(a, t) \exp[-i \bm{q}\cdot \bm{r}_i(0)],
\end{equation}
with $q = |\bm{q}|$.
Here $W_i(a, t) = \Theta(a - |\bm{r}_i(t) - \bm{r}_i(0)|)$ is the
overlap function or Heaviside step function $\Theta(x)$.
$W_i(a, t)$ selects the particle that moves farther than distance $a$
during the time interval $t$.
We use $a=0.3\sigma_1$ in this study.

The behavior of $S_4(q, t)$ at small wave numbers is conventionally described by
the Ornstein--Zernike (OZ) form as follows:
\begin{equation}
S_4(q, t) = \frac{\chi_4(t)}{1 + (q\xi_{4}(t))^{\alpha}},
\label{sk4_goz}
\end{equation}
where $\xi_4(t)$ is the correlation length and $\chi_4(t)$ is the
intensity at $q\to 0$.
As mentioned previously, it is an intricate task to numerically obtain
quantities such as the dynamical length scale from $S_4(q, t)$ in the $q
\to 0$ limit.
While our system size is smaller than optimal, we note that earlier work
in the same system with $N_1+N_2=100,000$ has presented extracted
lengths consistent with those we find here using the same method~\cite{Kim2013Multiple}.
Furthermore, the procedure we use to extract the dynamical length scale
$\xi_4(t)$, while not as rigorous is that used in
Refs~\onlinecite{Flenner2010Dynamic,Flenner2011Analysis}, has shown
consistency in extracted length values in the same system (compare the
results of Ref.~\onlinecite{Berthier2007Spontaneous} to those of
Ref.~\onlinecite{Karmakar2010Analysis})

\begin{figure*}[t]
\includegraphics[width=.9\textwidth]{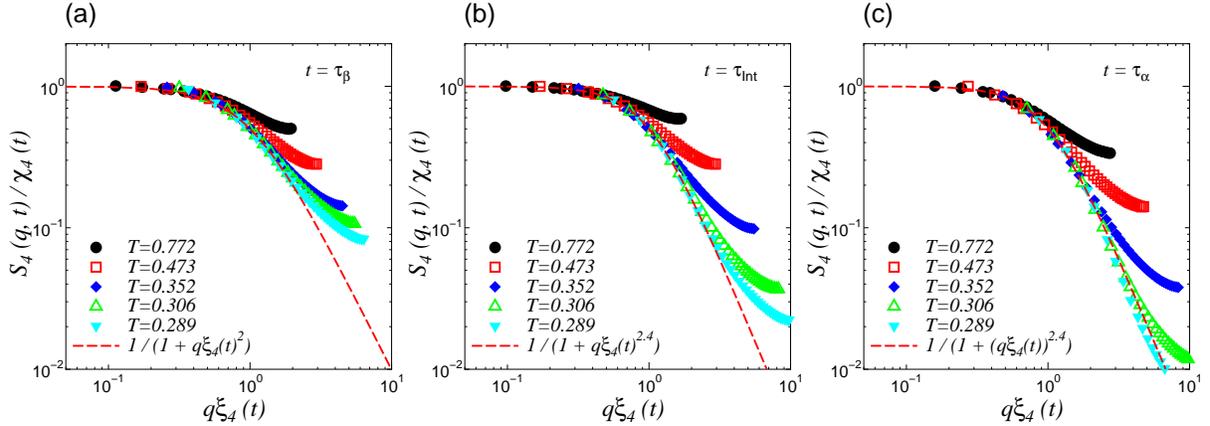}
\caption{
Scaled four-point correlation function $S_4(q, t) / \chi_4(t)$
as a function of $q\xi_4(t)$ at $T=0.772$, $0.473$, $0.352$, $0.306$,
and $0.289$.
The time interval $t$ is chosen as (a) $\tau_\beta$, (b) $\tau_{\rm Int}$, and (c) $\tau_\alpha$.
The dashed line represents the Ornstein--Zernike form $1/(1+(q\xi_4(t))^\alpha)$
with (a) $\alpha=2$, (b) $\alpha=2.4$, and (c) $\alpha=2.4$, respectively.
}
\label{sk4overlap_gozscale}
\end{figure*}

As shown in Fig.~\ref{sk4overlap_gozscale}(a), 
the results at the small time scale $t=\tau_{\beta}$ are well described
by Eq.~(\ref{sk4_goz}) with $\alpha=2$, which is the typical 
the OZ behaviour~\cite{Yamamoto1998Dynamics, Lacevic2003Spatially,
Berthier2004Time}.
In contrast, we find that for large time scales $t=\tau_{\rm Int}$ and
$\tau_\alpha$,
the slope of $S_4(q, t)$ becomes gradually sharper, which is more compatible
with a power $S_4(q, \tau_\alpha) \sim q^{-2.4}$ particularly at lower
temperatures, as demonstrated in Figs.~\ref{sk4overlap_gozscale}(b) and (c).
Similar power law behavior at $\alpha$-relaxation time has been reported
in the Kob--Andersen systems~\cite{Berthier2007Spontaneous}.
We also note that the same exponent $\alpha=2.4$ has been reported in the 
binary soft-sphere mixture with a larger system size
$N_1+N_2=100,000$~\cite{Kim2013Multiple}.
Thus, we choose $\alpha=2.4$ of Eq.~(\ref{sk4_goz}) and determine
$\xi_4(t)$ and $\chi_4(t)$
at two time scales $t =\tau_{\rm Int}$ and
$\tau_\alpha$ for various temperatures.

\begin{figure*}[t]
\includegraphics[width=.9\textwidth]{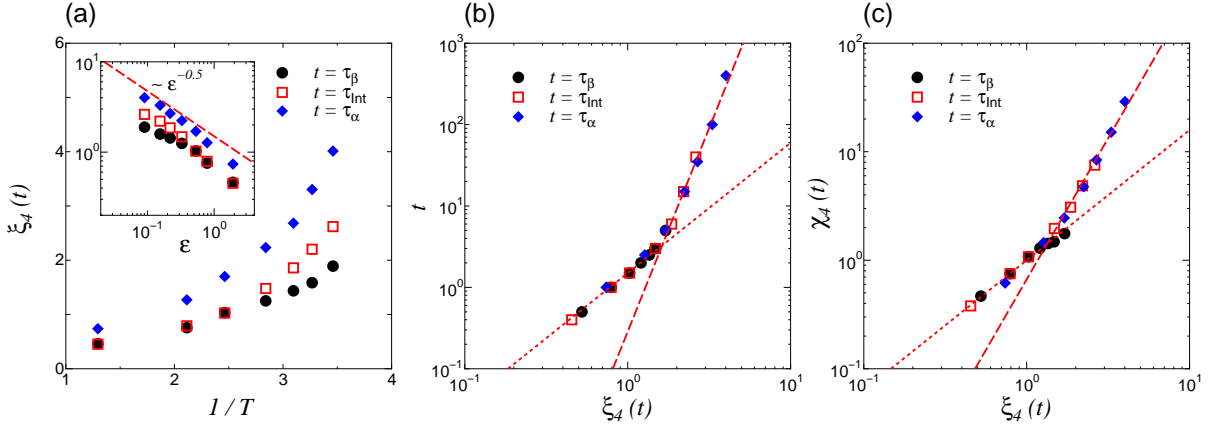}
\caption{
(a) Inverse temperature $1/T$ dependence of the length $\xi_4(t)$ at 
time scale $t =\{\tau_\beta, \tau_{\rm Int}, \tau_\alpha\}$.
Inset: $\xi_4$ as a function of $\epsilon=(T-T_c)/T_c$.
The dotted line represents the power law slope of $\xi_4 \sim \epsilon^{-0.5}$.
(b) Relationship between the time scale
$t =\{\tau_\beta, \tau_{\rm Int}, \tau_\alpha\}$ and length
$\xi_4(t)$ for various temperatures.
The dotted and dashed lines denotes the power law relationships of $t \sim
{\xi_4(t)}^z$ with $z=5$ and $z=1.6$, respectively.
(c) Relationship between the intensity $\chi_4(t)$ and
length $\xi_4(t)$ at time scale $t =\{\tau_\beta,
\tau_{\rm Int}, \tau_\alpha\}$ for various temperatures.
The dotted and dashed lines denote the power law relationships of $\chi_4(t) \sim
{\xi_4(t)}^{2-\eta}$ with $2-\eta =2.6$ and $2-\eta=1.2$, respectively.
}
\label{xi4}
\end{figure*}

The temperature dependence of the qualified length scale $\xi_4(t)$ 
at $t=\{\tau_\beta, \tau_{\rm Int}, \tau_\alpha\}$ is
shown in Fig~\ref{xi4}(a).
It is demonstrated that $\xi_4(t)$ at each time scale $t$
grows with decreasing temperature.
In particular, we observe the power law
behavior $\xi_4(\tau_\alpha) \sim \epsilon^{-\nu}$ with $\nu
\approx 0.5$ at the time scale of $\tau_\alpha$.
Note that IMCT instead predicts the exponent $\nu = 0.25$~\cite{Biroli2006Inhomogeneous}.
Thus, as found previously for a Kob--Andersen Lennard--Jones mixture, the IMCT results are
not compatible with the growth of
$\xi_4(\tau_\alpha)$~\cite{Berthier2007Spontaneous,
Karmakar2010Analysis, Kim2013Multiple}.
In addition, we examine the scaling relationships
between the time scale $t$ and length $\xi_4(t)$ and between the
intensity $\chi_4(t)$ and $\xi_4(t)$, which
are demonstrated in Figs.~\ref{xi4}(b) and (c), respectively.
The relationships are obtained as
$\tau_\alpha \sim
\xi_4(\tau_\alpha)^{z}$ with $z \approx \gamma/\nu \approx 5$ and 
$\chi_4(\tau_\alpha) \sim
\xi_4(\tau_\alpha)^{2-\eta}$ with $2-\eta \approx \alpha \approx 2.6$,
which are similar to those found in other systems. 
We also find the relationships at the
$\beta$-relaxation time regime with the smaller exponents, $z\approx 1.6$ and $2-\eta \approx 1.2$,
as observed in Figs.~\ref{xi4}(b) and (c).
At the intermediate time scale of $\tau_{\rm Int}$, the
cross-overs between two time scales, $\tau_\beta$ and $\tau_\alpha$,
are observed in those relationships.

\subsection{Three-point correlation function $\chi_u(k, q, t)$}
\label{Threepoint}

\begin{figure*}
\includegraphics[width=.6\textwidth]{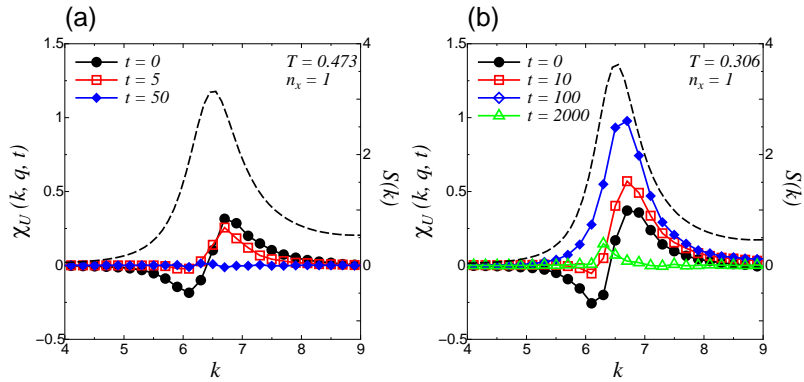}
\caption{
Wave number $k$ dependence of the three-point correlation function
$\chi_U(k, q, t)$ at (a) $T=0.473$ and (b) $T=0.306$ at
time interval $t$ using the left axis.
$n_x = 1$ corresponds to the smallest wave number $q_{\rm
  min}=2\pi/L \simeq 0.215$.
For comparison, the static structure factor $S(k)$ at each temperature
is plotted as a dashed line using the right axis.
}
\label{sk1k2t}
\end{figure*}

Here we present the numerical results of the three-point
correlation functions $\chi_U(k, q, t)$ obtained from the IMD simulations, 
as outlined in Sec.~\ref{NEMD}.
First, we show the wave number $k=|\bm{k}|$ dependence
of $\chi_U(k, q, t)$ for various time intervals $t$ in Fig.~\ref{sk1k2t}.
It is observed that at initial time $t=0$, the profile of $\chi_U(k, q, t)$ is
proportional to $\mathrm{d}S(k)/\mathrm{d}k$.
This property is reported in the mode-coupling calculation performed in
Ref.~\onlinecite{Szamel2010Diverging}.
At high temperature ($T=0.473$), the peak of $\chi_U(k, q, t)$
monotonically decreases as the time $t$ proceeds.
In contrast, at the supercooled state ($T=0.306$), $\chi_U(k, q,
t)$ develops a peak at the wave number where
$S(k)$ has its first peak, around $k\simeq 6.6$ when the time interval approaches
around the $\alpha$-relaxation, $\tau_\alpha (\simeq 100)$.
For larger times of $t \to \infty$, $\chi_U(k, q, t)$ tends to
decrease and finally becomes zero at any wave number $k$.

\begin{figure*}[t]
\includegraphics[width=.6\textwidth]{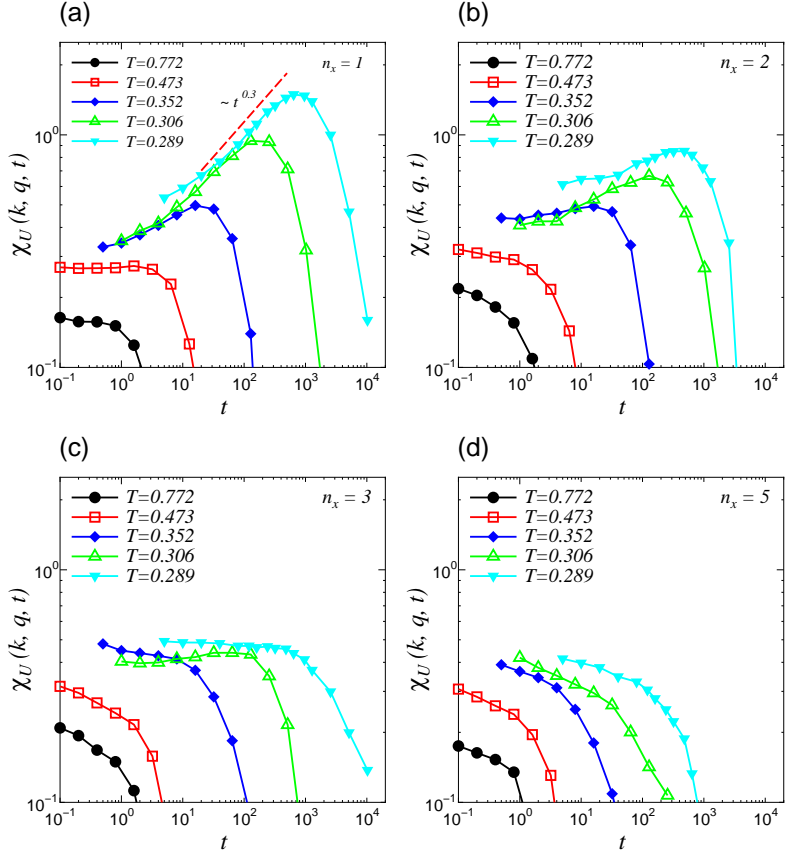}
\caption{
Time evolution of the three-point correlation function
$\chi_U(k, q, t)$ at $T=0.772$, $0.473$, $0.352$, $0.306$, and
$0.289$.
Wave number $q$ is chosen as $q=2\pi n_x/L$ with (a) $n_x=1$,
(b) $n_x=2$, (c) $n_x=3$, and (d) $n_x=5$.
T wave number $k$ is averaged over the range $k\in [6.5, 6.8]$.
}
\label{fk1k2t}
\end{figure*}

To observe how the three-point correlation function $\chi_U(k,
q, t)$ grows with time $t$, 
the time evolutions of $\chi_U(k, q, t)$ at various temperatures are
shown in Fig.~\ref{fk1k2t}.
Here the wave number $q=2\pi n_x/L$ of the external field is changed as
(a) $n_x=1$, (b) $n_x=2$, (c) $n_x=3$, and (d) $n_x=5$.
We averaged $\chi_U(\bm{k}, \bm{q}, t)$ over wave vectors $\bm{k}$
in the range of  $k\in [6.5, 6.8]$ to suppress statistical errors.
It is observed that the intensity of $\chi_U(k, q, t)$ with the
smallest wave number $n_x=1$ has its maximum value at 
the $\alpha$-relaxation time $\tau_\alpha$.
This basic feature of $\chi_U(k, q, t)$ is also demonstrated in
both the four-point susceptibilities $\chi_4(t)=\lim_{q\to 0} S_4(q, t)$
and as predicted by IMCT.
However, the time dependence of $\chi_U$ appears to grow as $t^{0.3}$ even at low
temperature, which is milder than that the growth $t^{0.7}$ we found for $\chi_4(t)$.
In addition, IMCT predicts $\chi_U \sim t^b$ with $b\approx 0.6$ in
the late $\beta$ regime and $t^a$ with $a\approx 0.3$ in the early $\beta$ regime.
The reason for this discrepancy between $\chi_4$ and IMCT is unclear. 
As already discussed 
the behavior of $\chi_4$ obtained as $q\rightarrow 0$
limit of $S_4$ is quite intricate because, roughly speaking, it is
expected to be given by two terms: one proportional to $\chi_U$ and
another proportional to its square.
The latter 
becoming important very close to $T_{MCT}$ but negligible far from
it~\cite{Berthier2007Spontaneous,
Berthier2007Spontaneous2}.
Thus obtaining reliable values of critical exponents 
from $\chi_4$ and $S_4$ is quite difficult. Another 
possibility is that the investigated
temperatures herein are still quite limited. As shown in
Fig.~\ref{msd}(a), even for the lowest temperature, the two-step
relaxation of the
intermediate scattering function is not well developed, making it difficult
to distinguish between the early and late $\beta$-relaxation regime.
Such limitations are imposed by the numerical difficulty in obtaining
well-averaged values of $\chi_U(k, q, t)$ from IMD.

\begin{figure*}[t]
\includegraphics[width=.9\textwidth]{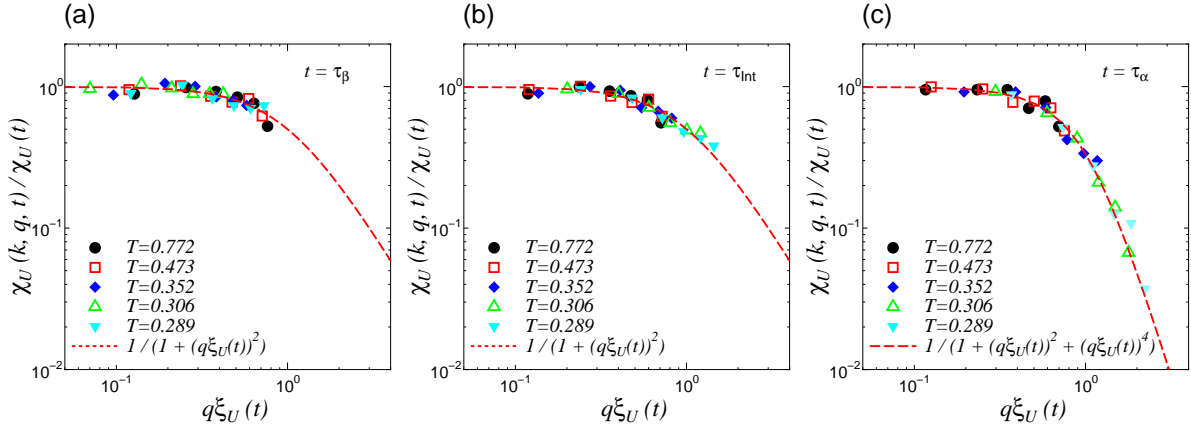}
\caption{
Scaled three-point correlation function 
$\chi_U(k, q, t)/ \chi_U(t)$ 
as a function of $q\xi_U(t)$ at $T=0.772$, $0.473$, $0.352$, $0.306$,
and $0.289$.
The time interval $t$ is chosen as (a) $\tau_\beta$, (b) $\tau_{\rm
 Int}$ (b), and (c) $\tau_\alpha$.
The dashed line represents the generalized OZ form
$1/(1 + (q\xi_{U}(t))^{2} + A((q\xi_{U}(t))^{4})$
with $A=0$ for (a) $t=\tau_\beta$ and (b) $\tau_{\rm Int}$, and (c) with
 $A=1$ for $t=\tau_\alpha$, respectively.
}
\label{chiu_scale}
\end{figure*}

Next, we examine the wave number $q$ dependence of the three-point
correlation function $\chi_U(k, q, t)$, which should be compared with
the four-point correlation function shown in Fig.~\ref{sk4overlap_gozscale}.
As shown in Fig.~\ref{fk1k2t}
the pronounced peak of $\chi_U$ rapidly decreases as the wave number
$q=2\pi n_x/L$ is increased, particularly at lower temperature.
To describe $\chi_U(k, q, t)$ and extract the
length scale, let us consider a generalized OZ form including a
$q^{-4}$ term as 
\begin{equation}
\chi_U(k, q, t) = \frac{\chi_U(t)}{1 + (q\xi_{U}(t))^{2} + A(q\xi_{U}(t))^{4}}.
\label{chiu_goz}
\end{equation}
Note that this $q^{-4}$ scaling is predicted by IMCT at the $\alpha$-relaxation time.
In the same way as Eq.~(\ref{sk4_goz}),
$\xi_U(t)$ and $\chi_U(t)$ express the length scale and intensity at $q\to 0$, respectively.
Figure~\ref{chiu_scale} shows
the scaled $\chi_U(k, q, t)/\chi_U(t)$ at various time scales
$t=\tau_\beta$, $\tau_{\rm Int}$, and $\tau_\alpha$.
First, as observed in Figs.~\ref{chiu_scale}(a) and (b),
$\chi_U(k, q, t)$ is well described by Eq.~(\ref{chiu_goz}) with $A=0$
at the time scales, $t=\tau_\beta$ and $\tau_{\rm Int}$,
corresponding to the usual OZ form.
In contrast, at the time scale $t=\tau_\alpha$, the wave number $q$ dependence of 
$\chi_U(k, q, t)$ becomes steeper than in $\chi_U(k, q, t)$ at smaller times
$t=\tau_\beta$ and $\tau_{\rm Int}$, as demonstrated in Fig.~\ref{chiu_scale}(c).
This behavior can be described by the expression in Eq.~(\ref{chiu_goz})
including the fourth order correction with $A=1$.
The observed cross-over of the function form $\chi_U$ 
from the vicinity of $\beta$-relaxation to $\alpha$-relaxation
is apparently different from the behavior of the four-point
correlations $S_4(q, t)$ observed in Fig.~\ref{sk4overlap_gozscale}; however, alternatively,
it is in accordance with the non-trivial prediction of IMCT~\cite{Biroli2006Inhomogeneous}.
Here we note that within IMCT the $q^{-4}$ scaling is not well developed
in the supercooled state ($\epsilon \simeq 0.1$) but instead becomes
clear much closer to $T_{MCT}$ (\textit{e.g.} $\epsilon \le
10^{-3}$)~\cite{Szamel2010Diverging}.
In this sense this distinction may well be an indicator of alteration of
the mean field behavior predicted by IMCT.

\begin{figure*}[t]
\includegraphics[width=.9\textwidth]{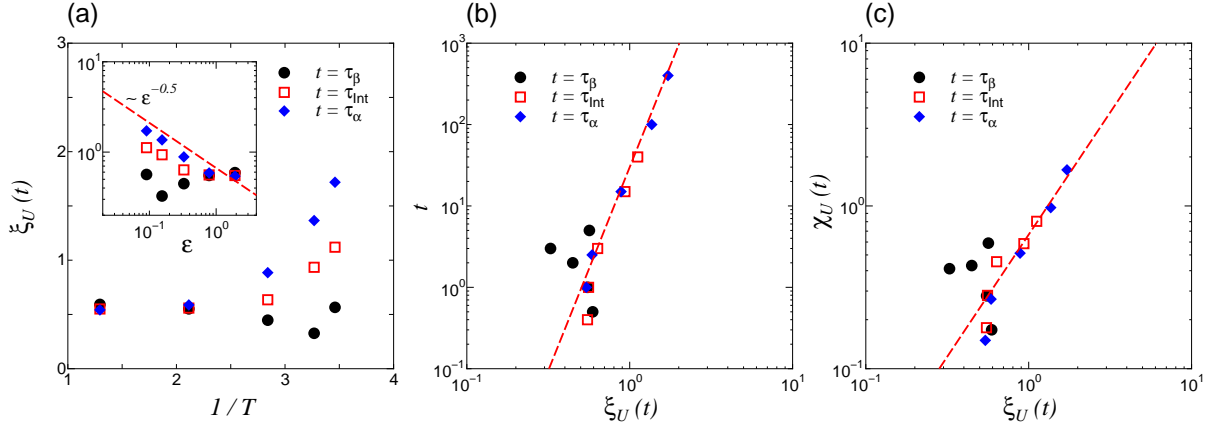}
\caption{
(a) Temperature dependence of $\xi_4(\tau_\alpha)$ and $\xi_U(\tau_\alpha)$.
Inset: $\xi_U(\tau_\alpha)$ as a function of $\epsilon=(T-T_c)/T_c$.
The dashed line is the power law slope of $\epsilon^{-0.5}$.
(b) Relationship between the time scale
$t =\{\tau_\beta, \tau_{\rm Int}, \tau_\alpha\}$ and length scale
$\xi_U(t)$ for various temperatures.
The dotted line is the power law relation, $t \sim {\xi_U}^{5}$.
(c) Relationship between the intensity $\chi_U(t)$ and
length $\xi_U(t)$ on the time scale $t =\{\tau_\beta,
\tau_{\rm Int}, \tau_\alpha\}$ for various temperatures.
The dashed line is the power law relationship, $\chi_U \sim {\xi_U}^{1.5}$.
}
\label{xiu}
\end{figure*}

The determined length scale $\xi_U(t)$ is shown 
as a function of temperature $T$ in Fig.~\ref{xiu}(a) at times,
$t={\tau_\beta}$, $\tau_{\rm Int}$, and $\tau_{\alpha}$.
Similar to the temperature dependence of the length scale $\xi_4$
extracted from the four-point correlator,
the length scale $\xi_U(\tau_\alpha)$ increases with decreasing temperature.
Although the evaluated value $\xi_U(\tau_\alpha)$ is smaller than $\xi_4(\tau_\alpha)$,
$\xi_U(\tau_\alpha)$ can be approximated by $\xi_U(\tau_\alpha)\sim \epsilon
^{-\nu}$ with $\nu\approx 0.5$, which is same as the $\xi_4(\tau_\alpha)$ (see Fig.~\ref{xi4}(a)).
Since absolute length scales are not obtained via the scaling analysis
performed here, the agreement in scaling between $\xi_U(\tau_\alpha)$ and $\xi_4(\tau_\alpha)$ should
be taken as preliminary confirmation of the generic prediction from the
analysis of Ref.~\onlinecite{Berthier2007Spontaneous}
that $\xi_U \sim \xi_4$.  Furthermore, the relationship $\tau_\alpha \sim
\xi_U(\tau_\alpha)^z$ with $z \approx 5$ is observed in Fig.~\ref{xiu}(b).
This exponent is close to the value of
$\xi_4(\tau_\alpha)$, as obtained in Fig.~\ref{xi4}(b).
We also obtain the relationship $\chi_U(\tau_\alpha)\sim
\xi_U(\tau_\alpha)^{2-\eta_\alpha}$ with $2-\eta_\alpha \approx 1.5$
in Fig.~\ref{xiu}(c).
This exponent is rather smaller than that of $\xi_4(\tau_\alpha)$ obtained in Fig.~\ref{xi4}(c).
A disagreement with the IMCT prediction $2-\eta_\alpha = 4$ is also observed.
In addition,
as shown in Fig.~\ref{xiu}(a), the length scale $\xi_U(\tau_\beta)$ at
$\tau_\beta$ is not available because of the large numerical fluctuations.
Here it can be considered that the minimum wave number of the present system $q_{\rm min}=2\pi/L
\simeq 0.215$ is still too large to reduce those numerical errors.
As mentioned above, it is an important future goal to seek  lower temperature data approaching
the mode-coupling transition temperature.
In particular, further analysis for larger systems and lower temperatures  is necessary
to improve the signal-to-noise ratio of the response function and acquire
more insight into the behavior of $\chi_U$.

\begin{figure*}[t]
\includegraphics[width=.3\textwidth]{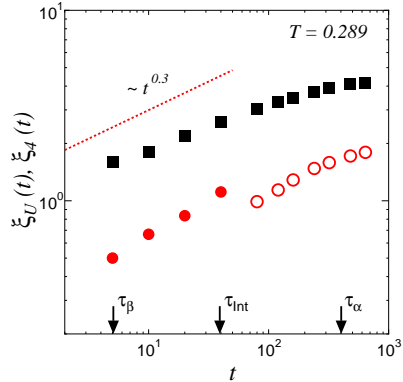}
\caption{
Time dependence of the length scales $\xi_4(t)$ (squares) and $\xi_U(t)$
 (circles) at the lowest temperature $T=0.289$.
The time values $\tau_\alpha$, $\tau_{\rm Int}$, and $\tau_\beta$ are
 indicated by arrows.
Note that for the length scale $\xi_U(t)$, the closed and open symbols denote
$\xi_U(t)$ evaluated by the generalized OZ form Eq.~(\ref{chiu_goz})
 with $A=0$ and $A=1$, respectively.
This switch is responsible for the small gap of $\xi_U$ around $\tau_{\rm Int}$.
}
\label{xi147}
\end{figure*}

Finally, the physical implementation of the time-scale dependence of the
function form $\chi_U$ is worthy of mention.
As discussed in Ref.~\onlinecite{Biroli2006Inhomogeneous}, 
this cross-over of the scaling function might be relevant to the
geometrical change of dynamically correlated motions.
Namely, IMCT predicts that the dynamic length scale $\xi_U(t)$ increases at
the early $\beta$
regime and then saturates to a constant value at the late $\beta$ regime.
This suggests that while keeping a constant spatial extent, the geometrical structure of the dynamic
correlations significantly fatten between $\tau_\beta$ and $\tau_\alpha$.
Recent MD simulations reveal that the mobile particle motions
form string-like structures in the $\beta$-relaxation
regime~\cite{Donati1998Stringlike}, whereas a
more compact structure is observed at the slower time scale of
$\tau_\alpha$~\cite{Appignanesi2006Democratic}.
In Fig.~\ref{xi147}, we show the time evolution of the length
scales $\xi_4(t)$ and $\xi_U(t)$ at the lowest temperature $T=0.289$.
Although it is difficult to distinguish between early and late $\beta$ regimes in
the present simulation, we observe that both length scales tend to
increase from $\tau_\beta$ and saturate around the time exceeding
$\tau_{\rm Int}$ in a similar manner.
After the time scale $\tau_{\rm Int}$ a better description of $\chi_U$ is obtained with the fourth-order
corrections in the generalized OZ form, Eq.~(\ref{chiu_goz}), as
shown in Fig.~\ref{chiu_scale}(c).

\section{Summary and Conclusions}
\label{Summary}

In Refs.~\onlinecite{ Berthier2007Spontaneous,
Berthier2007Spontaneous2} it was argued on general grounds, beyond the particularities of mean-field predictions
that originate from theories such as IMCT, that the function $\chi_U(\bm{k}, \bm{q}, t)$, which 
 is the response of the two-point correlation function with respect to an inhomogeneous external field, offers
particular advantages over the more conventional $S_4(q, t)$. 
For these reasons we have investigated the function
 $\chi_U(\bm{k}, \bm{q}, t)$ to quantitatively characterize the length scale of
dynamic heterogeneity via IMD simulations.
As predicted by IMCT, we did find that $\xi_U$ probe dynamic correlations and that the
associated dynamic correlation length scales similarly to the one extracted from $S_4$. 
Therefore $\xi_U$ provides a viable alternative to $\chi_4$ and $S_4$: its advantages are 
an enhanced possibility for experimental extraction in colloidal systems, as well as 
a simpler ensemble and temperature dependence.
Thus extracting critical properties of
$\chi_4$ is quite delicate.

We also compared the critical behavior of $\chi_U$ obtained in
simulations to IMCT results. 
Although some predictions are verified, as
the cross-over 
from the $q^{-2}$ decay to the sharper $q^{-4}$ decay when the time
scale changes from $\tau_\beta$ to $\tau_\alpha$, others are not. For
example the values of $\nu$ and $z_\alpha,z_\beta$
are off by a substantial amount. 
Moreover, there is a discrepancy 
between the value of $\eta_\alpha$ obtained from the $q^{-4}$ decay, 
which is in agreement with IMCT, and the one obtained by 
the relation $\chi_4(\tau_\alpha)\propto
\xi_U(\tau_\alpha)^{2-\eta_\alpha}$.
This difference could either indicate a breakdown of usual scaling laws or, more simply,
that the systems is not close enough to the critical point and, hence,
there is a substantial error in the values of the exponents. 
We have indications that the latter option is the most likely one. 
Further analysis is necessary to
assess the behavior of  $\chi_U(\bm{k}, \bm{q}, t)$ more critically in a 
much larger and varied systems and at a lower temperatures.

Finally, it is important to recall that 
the IMCT exponents are mean-field ones and that the upper critical dimension for the
MCT transition is $d_u=8$~\cite{Biroli2007Critical, Franz2011Field}.
In fact, as previously mentioned, in three dimensions non-mean field
fluctuations such thermally activated hopping motion occurs.
In consequence, the fact that IMCT works qualitatively 
but not quantitatively is actually a promising evidence that dynamic
correlations close to $T_{MCT}$ can be indeed described in terms of a
dynamical critical MCT phenomenon but that in order to reach
quantitative agreement a theory of critical fluctuations valid below
$d_u$ has to be constructed.
Progress in this direction have been
recently obtained in Refs. \onlinecite{Franz2011Field, Franz2010Properties}.

\begin{acknowledgement}

We thank J.-P. Bouchaud for discussions and collaboration on the topics addressed
in this work, in particular IMCT. 
K.K. was supported by Grants-in-Aid for Scientific Research:
Young Scientists (A) No. 23684037 from Japan Society for the Promotion of Science (JSPS).
D.R.R. was supported by NSF CHE-1213247.
G.B. was supported by the ERC grant NPRGGLASS.
K.K. acknowledges Yamada Science Foundation for supporting his stay at
Columbia University in 2011.
The computations were performed at Research Center for Computational
Science, Okazaki Research Facilities, National Institutes of Natural Sciences, Japan.

\end{acknowledgement}

%
%


\begin{mcitethebibliography}{49}
\providecommand*\natexlab[1]{#1}
\providecommand*\mciteSetBstSublistMode[1]{}
\providecommand*\mciteSetBstMaxWidthForm[2]{}
\providecommand*\mciteBstWouldAddEndPuncttrue
  {\def\EndOfBibitem{\unskip.}}
\providecommand*\mciteBstWouldAddEndPunctfalse
  {\let\EndOfBibitem\relax}
\providecommand*\mciteSetBstMidEndSepPunct[3]{}
\providecommand*\mciteSetBstSublistLabelBeginEnd[3]{}
\providecommand*\EndOfBibitem{}
\mciteSetBstSublistMode{f}
\mciteSetBstMaxWidthForm{subitem}{(\alph{mcitesubitemcount})}
\mciteSetBstSublistLabelBeginEnd
  {\mcitemaxwidthsubitemform\space}
  {\relax}
  {\relax}

\bibitem[Wolynes and Lubchenko(2012)Wolynes, and
  Lubchenko]{Wolynes2012Structural}
Wolynes,~P.~G.; Lubchenko,~V. \emph{{Structural Glasses and Supercooled
  Liquids}}; John Wiley \& Sons, USA, 2012\relax
\mciteBstWouldAddEndPuncttrue
\mciteSetBstMidEndSepPunct{\mcitedefaultmidpunct}
{\mcitedefaultendpunct}{\mcitedefaultseppunct}\relax
\EndOfBibitem
\bibitem[Ediger et~al.(1996)Ediger, Angell, and Nagel]{Ediger1996Supercooled}
Ediger,~M.~D.; Angell,~C.~A.; Nagel,~S.~R. {Supercooled Liquids and Glasses}.
  \emph{J. Phys. Chem.} \textbf{1996}, \emph{100}, 13200--13212\relax
\mciteBstWouldAddEndPuncttrue
\mciteSetBstMidEndSepPunct{\mcitedefaultmidpunct}
{\mcitedefaultendpunct}{\mcitedefaultseppunct}\relax
\EndOfBibitem
\bibitem[Debenedetti and Stillinger(2001)Debenedetti, and
  Stillinger]{Debenedetti2001Supercooled}
Debenedetti,~P.~G.; Stillinger,~F.~H. {Supercooled liquids and the glass
  transition}. \emph{Nature} \textbf{2001}, \emph{410}, 259--267\relax
\mciteBstWouldAddEndPuncttrue
\mciteSetBstMidEndSepPunct{\mcitedefaultmidpunct}
{\mcitedefaultendpunct}{\mcitedefaultseppunct}\relax
\EndOfBibitem
\bibitem[Lubchenko and Wolynes(2007)Lubchenko, and
  Wolynes]{Lubchenko2007Theory}
Lubchenko,~V.; Wolynes,~P.~G. {Theory of Structural Glasses and Supercooled
  Liquids}. \emph{Annu. Rev. Phys. Chem.} \textbf{2007}, \emph{58},
  235--266\relax
\mciteBstWouldAddEndPuncttrue
\mciteSetBstMidEndSepPunct{\mcitedefaultmidpunct}
{\mcitedefaultendpunct}{\mcitedefaultseppunct}\relax
\EndOfBibitem
\bibitem[Cavagna(2009)]{Cavagna2009Supercooled}
Cavagna,~A. {Supercooled liquids for pedestrians}. \emph{Phys. Rep.}
  \textbf{2009}, \emph{476}, 51--124\relax
\mciteBstWouldAddEndPuncttrue
\mciteSetBstMidEndSepPunct{\mcitedefaultmidpunct}
{\mcitedefaultendpunct}{\mcitedefaultseppunct}\relax
\EndOfBibitem
\bibitem[Berthier and Biroli(2011)Berthier, and
  Biroli]{Berthier2011Theoretical}
Berthier,~L.; Biroli,~G. {Theoretical perspective on the glass transition and
  amorphous materials}. \emph{Rev. Mod. Phys.} \textbf{2011}, \emph{83},
  587--645\relax
\mciteBstWouldAddEndPuncttrue
\mciteSetBstMidEndSepPunct{\mcitedefaultmidpunct}
{\mcitedefaultendpunct}{\mcitedefaultseppunct}\relax
\EndOfBibitem
\bibitem[Berthier et~al.(2011)Berthier, Biroli, Bouchaud, Cipelletti, and van
  Saarloos]{Berthier2011Dynamical}
Berthier,~L., Biroli,~G., Bouchaud,~J.-P., Cipelletti,~L., van Saarloos,~W.,
  Eds. \emph{{Dynamical Heterogeneities in Glasses, Colloids, and Granular
  Media}}; Oxford University Press: USA, 2011\relax
\mciteBstWouldAddEndPuncttrue
\mciteSetBstMidEndSepPunct{\mcitedefaultmidpunct}
{\mcitedefaultendpunct}{\mcitedefaultseppunct}\relax
\EndOfBibitem
\bibitem[Ediger(2000)]{Ediger2000Spatially}
Ediger,~M.~D. {Spatially heterogeneous dynamics in supercooled liquids.}
  \emph{Annu. Rev. Phys. Chem.} \textbf{2000}, \emph{51}, 99--128\relax
\mciteBstWouldAddEndPuncttrue
\mciteSetBstMidEndSepPunct{\mcitedefaultmidpunct}
{\mcitedefaultendpunct}{\mcitedefaultseppunct}\relax
\EndOfBibitem
\bibitem[Hurley and Harrowell(1995)Hurley, and Harrowell]{Hurley1995Kinetic}
Hurley,~M.~M.; Harrowell,~P. {Kinetic structure of a two-dimensional liquid}.
  \emph{Phys. Rev. E} \textbf{1995}, \emph{52}, 1694--1698\relax
\mciteBstWouldAddEndPuncttrue
\mciteSetBstMidEndSepPunct{\mcitedefaultmidpunct}
{\mcitedefaultendpunct}{\mcitedefaultseppunct}\relax
\EndOfBibitem
\bibitem[Kob et~al.(1997)Kob, Donati, Plimpton, Poole, and
  Glotzer]{Kob1997Dynamical}
Kob,~W.; Donati,~C.; Plimpton,~S.~J.; Poole,~P.~H.; Glotzer,~S.~C. {Dynamical
  Heterogeneities in a Supercooled Lennard-Jones Liquid}. \emph{Phys. Rev.
  Lett.} \textbf{1997}, \emph{79}, 2827\relax
\mciteBstWouldAddEndPuncttrue
\mciteSetBstMidEndSepPunct{\mcitedefaultmidpunct}
{\mcitedefaultendpunct}{\mcitedefaultseppunct}\relax
\EndOfBibitem
\bibitem[Yamamoto and Onuki(1997)Yamamoto, and Onuki]{Yamamoto1997Kinetic}
Yamamoto,~R.; Onuki,~A. {Kinetic Heterogeneities in a Highly Supercooled
  Liquid}. \emph{J. Phys. Soc. Jpn.} \textbf{1997}, \emph{66}, 2545--2548\relax
\mciteBstWouldAddEndPuncttrue
\mciteSetBstMidEndSepPunct{\mcitedefaultmidpunct}
{\mcitedefaultendpunct}{\mcitedefaultseppunct}\relax
\EndOfBibitem
\bibitem[Yamamoto and Onuki(1998)Yamamoto, and Onuki]{Yamamoto1998Dynamics}
Yamamoto,~R.; Onuki,~A. {Dynamics of highly supercooled liquids: Heterogeneity,
  rheology, and diffusion}. \emph{Phys. Rev. E} \textbf{1998}, \emph{58},
  3515--3529\relax
\mciteBstWouldAddEndPuncttrue
\mciteSetBstMidEndSepPunct{\mcitedefaultmidpunct}
{\mcitedefaultendpunct}{\mcitedefaultseppunct}\relax
\EndOfBibitem
\bibitem[Franz and Parisi(2000)Franz, and Parisi]{Franz2000On}
Franz,~S.; Parisi,~G. {On non-linear susceptibility in supercooled liquids}.
  \emph{J. Phys.: Condens. Matter} \textbf{2000}, \emph{12}, 6335--6342\relax
\mciteBstWouldAddEndPuncttrue
\mciteSetBstMidEndSepPunct{\mcitedefaultmidpunct}
{\mcitedefaultendpunct}{\mcitedefaultseppunct}\relax
\EndOfBibitem
\bibitem[Donati et~al.(2002)Donati, Franz, Glotzer, and
  Parisi]{Donati2002Theory}
Donati,~C.; Franz,~S.; Glotzer,~S.~C.; Parisi,~G. {Theory of non-linear
  susceptibility and correlation length in glasses and liquids}. \emph{J.
  Non-Cryst. Solids} \textbf{2002}, \emph{307-310}, 215--224\relax
\mciteBstWouldAddEndPuncttrue
\mciteSetBstMidEndSepPunct{\mcitedefaultmidpunct}
{\mcitedefaultendpunct}{\mcitedefaultseppunct}\relax
\EndOfBibitem
\bibitem[La{\v c}evi{\'c} et~al.(2002)La{\v c}evi{\'c}, Starr, Schr{\o}der,
  Novikov, and Glotzer]{Lacevic2002Growing}
La{\v c}evi{\'c},~N.; Starr,~F.; Schr{\o}der,~T.; Novikov,~V.; Glotzer,~S.
  {Growing correlation length on cooling below the onset of caging in a
  simulated glass-forming liquid}. \emph{Phys. Rev. E} \textbf{2002},
  \emph{66}, 030101\relax
\mciteBstWouldAddEndPuncttrue
\mciteSetBstMidEndSepPunct{\mcitedefaultmidpunct}
{\mcitedefaultendpunct}{\mcitedefaultseppunct}\relax
\EndOfBibitem
\bibitem[La{\v c}evi{\'c} et~al.(2003)La{\v c}evi{\'c}, Starr, Schr{\o}der, and
  Glotzer]{Lacevic2003Spatially}
La{\v c}evi{\'c},~N.; Starr,~F.~W.; Schr{\o}der,~T.~B.; Glotzer,~S.~C.
  {Spatially heterogeneous dynamics investigated via a time-dependent
  four-point density correlation function}. \emph{J. Chem. Phys.}
  \textbf{2003}, \emph{119}, 7372--7387\relax
\mciteBstWouldAddEndPuncttrue
\mciteSetBstMidEndSepPunct{\mcitedefaultmidpunct}
{\mcitedefaultendpunct}{\mcitedefaultseppunct}\relax
\EndOfBibitem
\bibitem[Berthier(2004)]{Berthier2004Time}
Berthier,~L. {Time and length scales in supercooled liquids}. \emph{Phys. Rev.
  E} \textbf{2004}, \emph{69}, 020201(R)\relax
\mciteBstWouldAddEndPuncttrue
\mciteSetBstMidEndSepPunct{\mcitedefaultmidpunct}
{\mcitedefaultendpunct}{\mcitedefaultseppunct}\relax
\EndOfBibitem
\bibitem[Whitelam et~al.(2004)Whitelam, Berthier, and
  Garrahan]{Whitelam2004Dynamic}
Whitelam,~S.; Berthier,~L.; Garrahan,~J.~P. {Dynamic Criticality in
  Glass-Forming Liquids}. \emph{Phys. Rev. Lett.} \textbf{2004}, \emph{92},
  185705\relax
\mciteBstWouldAddEndPuncttrue
\mciteSetBstMidEndSepPunct{\mcitedefaultmidpunct}
{\mcitedefaultendpunct}{\mcitedefaultseppunct}\relax
\EndOfBibitem
\bibitem[Toninelli et~al.(2005)Toninelli, Wyart, Berthier, Biroli, and
  Bouchaud]{Toninelli2005Dynamical}
Toninelli,~C.; Wyart,~M.; Berthier,~L.; Biroli,~G.; Bouchaud,~J.~P. {Dynamical
  susceptibility of glass formers: Contrasting the predictions of theoretical
  scenarios}. \emph{Phys. Rev. E} \textbf{2005}, \emph{71}, 041505\relax
\mciteBstWouldAddEndPuncttrue
\mciteSetBstMidEndSepPunct{\mcitedefaultmidpunct}
{\mcitedefaultendpunct}{\mcitedefaultseppunct}\relax
\EndOfBibitem
\bibitem[Chandler et~al.(2006)Chandler, Garrahan, Jack, Maibaum, and
  Pan]{Chandler2006Lengthscale}
Chandler,~D.; Garrahan,~J.~P.; Jack,~R.~L.; Maibaum,~L.; Pan,~A.~C.
  {Lengthscale dependence of dynamic four-point susceptibilities in glass
  formers}. \emph{Phys. Rev. E} \textbf{2006}, \emph{74}, 051501\relax
\mciteBstWouldAddEndPuncttrue
\mciteSetBstMidEndSepPunct{\mcitedefaultmidpunct}
{\mcitedefaultendpunct}{\mcitedefaultseppunct}\relax
\EndOfBibitem
\bibitem[Szamel and Flenner(2006)Szamel, and Flenner]{Szamel2006Four}
Szamel,~G.; Flenner,~E. {Four-point susceptibility of a glass-forming binary
  mixture: Brownian dynamics}. \emph{Phys. Rev. E} \textbf{2006}, \emph{74},
  021507\relax
\mciteBstWouldAddEndPuncttrue
\mciteSetBstMidEndSepPunct{\mcitedefaultmidpunct}
{\mcitedefaultendpunct}{\mcitedefaultseppunct}\relax
\EndOfBibitem
\bibitem[Berthier et~al.(2007)Berthier, Biroli, Bouchaud, Kob, Miyazaki, and
  Reichman]{Berthier2007Spontaneous}
Berthier,~L. et~al.  {Spontaneous and induced dynamic fluctuations in glass
  formers. I. General results and dependence on ensemble and dynamics}.
  \emph{J. Chem. Phys.} \textbf{2007}, \emph{126}, 184503\relax
\mciteBstWouldAddEndPuncttrue
\mciteSetBstMidEndSepPunct{\mcitedefaultmidpunct}
{\mcitedefaultendpunct}{\mcitedefaultseppunct}\relax
\EndOfBibitem
\bibitem[Berthier et~al.(2007)Berthier, Biroli, Bouchaud, Kob, Miyazaki, and
  Reichman]{Berthier2007Spontaneous2}
Berthier,~L. et~al.  {Spontaneous and induced dynamic correlations in glass
  formers. II. Model calculations and comparison to numerical simulations}.
  \emph{J. Chem. Phys.} \textbf{2007}, \emph{126}, 184504\relax
\mciteBstWouldAddEndPuncttrue
\mciteSetBstMidEndSepPunct{\mcitedefaultmidpunct}
{\mcitedefaultendpunct}{\mcitedefaultseppunct}\relax
\EndOfBibitem
\bibitem[Stein and Andersen(2008)Stein, and Andersen]{Stein2008Scaling}
Stein,~R. S.~L.; Andersen,~H.~C. {Scaling Analysis of Dynamic Heterogeneity in
  a Supercooled Lennard-Jones Liquid}. \emph{Phys. Rev. Lett.} \textbf{2008},
  \emph{101}, 267802\relax
\mciteBstWouldAddEndPuncttrue
\mciteSetBstMidEndSepPunct{\mcitedefaultmidpunct}
{\mcitedefaultendpunct}{\mcitedefaultseppunct}\relax
\EndOfBibitem
\bibitem[Karmakar et~al.(2009)Karmakar, Dasgupta, and
  Sastry]{Karmakar2009Growing}
Karmakar,~S.; Dasgupta,~C.; Sastry,~S. {Growing length and time scales in
  glass-forming liquids}. \emph{Proc. Natl. Acad. Sci. U.S.A.} \textbf{2009},
  \emph{106}, 3675--3679\relax
\mciteBstWouldAddEndPuncttrue
\mciteSetBstMidEndSepPunct{\mcitedefaultmidpunct}
{\mcitedefaultendpunct}{\mcitedefaultseppunct}\relax
\EndOfBibitem
\bibitem[Karmakar et~al.(2010)Karmakar, Dasgupta, and
  Sastry]{Karmakar2010Analysis}
Karmakar,~S.; Dasgupta,~C.; Sastry,~S. {Analysis of Dynamic Heterogeneity in a
  Glass Former from the Spatial Correlations of Mobility}. \emph{Phys. Rev.
  Lett.} \textbf{2010}, \emph{105}, 015701\relax
\mciteBstWouldAddEndPuncttrue
\mciteSetBstMidEndSepPunct{\mcitedefaultmidpunct}
{\mcitedefaultendpunct}{\mcitedefaultseppunct}\relax
\EndOfBibitem
\bibitem[Flenner and Szamel(2010)Flenner, and Szamel]{Flenner2010Dynamic}
Flenner,~E.; Szamel,~G. {Dynamic Heterogeneity in a Glass Forming Fluid:
  Susceptibility, Structure Factor, and Correlation Length}. \emph{Phys. Rev.
  Lett.} \textbf{2010}, \emph{105}, 217801\relax
\mciteBstWouldAddEndPuncttrue
\mciteSetBstMidEndSepPunct{\mcitedefaultmidpunct}
{\mcitedefaultendpunct}{\mcitedefaultseppunct}\relax
\EndOfBibitem
\bibitem[Flenner et~al.(2011)Flenner, Zhang, and Szamel]{Flenner2011Analysis}
Flenner,~E.; Zhang,~M.; Szamel,~G. {Analysis of a growing dynamic length scale
  in a glass-forming binary hard-sphere mixture}. \emph{Phys. Rev. E}
  \textbf{2011}, \emph{83}, 051501\relax
\mciteBstWouldAddEndPuncttrue
\mciteSetBstMidEndSepPunct{\mcitedefaultmidpunct}
{\mcitedefaultendpunct}{\mcitedefaultseppunct}\relax
\EndOfBibitem
\bibitem[Mizuno and Yamamoto(2011)Mizuno, and Yamamoto]{Mizuno2011Dynamical}
Mizuno,~H.; Yamamoto,~R. {Dynamical heterogeneity in a highly supercooled
  liquid: Consistent calculations of correlation length, intensity, and
  lifetime}. \emph{Phys. Rev. E} \textbf{2011}, \emph{84}, 011506\relax
\mciteBstWouldAddEndPuncttrue
\mciteSetBstMidEndSepPunct{\mcitedefaultmidpunct}
{\mcitedefaultendpunct}{\mcitedefaultseppunct}\relax
\EndOfBibitem
\bibitem[Kim and Saito(2013)Kim, and Saito]{Kim2013Multiple}
Kim,~K.; Saito,~S. {Multiple length and time scales of dynamic heterogeneities
  in model glass-forming liquids: A systematic analysis of multi-point and
  multi-time correlations}. \emph{J. Chem. Phys.} \textbf{2013}, \emph{138},
  12A506\relax
\mciteBstWouldAddEndPuncttrue
\mciteSetBstMidEndSepPunct{\mcitedefaultmidpunct}
{\mcitedefaultendpunct}{\mcitedefaultseppunct}\relax
\EndOfBibitem
\bibitem[Berthier et~al.(2005)Berthier, Biroli, Bouchaud, Cipelletti, Masri,
  L'H\^{o}te, Ladieu, and Pierno]{Berthier2005Direct}
Berthier,~L. et~al.  {Direct Experimental Evidence of a Growing Length Scale
  Accompanying the Glass Transition}. \emph{Science} \textbf{2005}, \emph{310},
  1797--1800\relax
\mciteBstWouldAddEndPuncttrue
\mciteSetBstMidEndSepPunct{\mcitedefaultmidpunct}
{\mcitedefaultendpunct}{\mcitedefaultseppunct}\relax
\EndOfBibitem
\bibitem[Dalle-Ferrier et~al.(2007)Dalle-Ferrier, Thibierge, Alba-Simionesco,
  Berthier, Biroli, Bouchaud, Ladieu, L'H\^{o}te, and
  Tarjus]{DalleFirrier2007Spatial}
Dalle-Ferrier,~C. et~al.  {Spatial correlations in the dynamics of glassforming
  liquids: Experimental determination of their temperature dependence}.
  \emph{Phys. Rev. E} \textbf{2007}, \emph{76}, 041510\relax
\mciteBstWouldAddEndPuncttrue
\mciteSetBstMidEndSepPunct{\mcitedefaultmidpunct}
{\mcitedefaultendpunct}{\mcitedefaultseppunct}\relax
\EndOfBibitem
\bibitem[Brambilla et~al.(2009)Brambilla, Masri, Pierno, Berthier, Cipelletti,
  Petekidis, and Schofield]{Brambilla2009Probing}
Brambilla,~G. et~al.  {Probing the Equilibrium Dynamics of Colloidal Hard
  Spheres above the Mode-Coupling Glass Transition}. \emph{Phys. Rev. Lett.}
  \textbf{2009}, \emph{102}, 085703\relax
\mciteBstWouldAddEndPuncttrue
\mciteSetBstMidEndSepPunct{\mcitedefaultmidpunct}
{\mcitedefaultendpunct}{\mcitedefaultseppunct}\relax
\EndOfBibitem
\bibitem[Bouchaud and Biroli(2005)Bouchaud, and Biroli]{Bouchaud2005Nonlinear}
Bouchaud,~J.~P.; Biroli,~G. {Nonlinear susceptibility in glassy systems: A
  probe for cooperative dynamical length scales}. \emph{Phys. Rev. B}
  \textbf{2005}, \emph{72}, 064204\relax
\mciteBstWouldAddEndPuncttrue
\mciteSetBstMidEndSepPunct{\mcitedefaultmidpunct}
{\mcitedefaultendpunct}{\mcitedefaultseppunct}\relax
\EndOfBibitem
\bibitem[Tarzia et~al.(2010)Tarzia, Biroli, Lef\`{e}vre, and
  Bouchaud]{Tarzia2010Anomalous}
Tarzia,~M.; Biroli,~G.; Lef\`{e}vre,~A.; Bouchaud,~J.~P. {Anomalous nonlinear
  response of glassy liquids: General arguments and a mode-coupling approach}.
  \emph{J. Chem. Phys.} \textbf{2010}, \emph{132}, 054501\relax
\mciteBstWouldAddEndPuncttrue
\mciteSetBstMidEndSepPunct{\mcitedefaultmidpunct}
{\mcitedefaultendpunct}{\mcitedefaultseppunct}\relax
\EndOfBibitem
\bibitem[Crauste-Thibierge et~al.(2010)Crauste-Thibierge, Brun, Ladieu,
  L'H\^{o}te, Biroli, and Bouchaud]{CrausteThibierge2010Evidence}
Crauste-Thibierge,~C. et~al.  {Evidence of Growing Spatial Correlations at the
  Glass Transition from Nonlinear Response Experiments}. \emph{Phys. Rev.
  Lett.} \textbf{2010}, \emph{104}, 165703\relax
\mciteBstWouldAddEndPuncttrue
\mciteSetBstMidEndSepPunct{\mcitedefaultmidpunct}
{\mcitedefaultendpunct}{\mcitedefaultseppunct}\relax
\EndOfBibitem
\bibitem[Diezemann(2012)]{Diezemann2012Nonlinear}
Diezemann,~G. {Nonlinear response theory for Markov processes: Simple models
  for glassy relaxation}. \emph{Phys. Rev. E} \textbf{2012}, \emph{85},
  051502\relax
\mciteBstWouldAddEndPuncttrue
\mciteSetBstMidEndSepPunct{\mcitedefaultmidpunct}
{\mcitedefaultendpunct}{\mcitedefaultseppunct}\relax
\EndOfBibitem
\bibitem[Biroli et~al.(2006)Biroli, Bouchaud, Miyazaki, and
  Reichman]{Biroli2006Inhomogeneous}
Biroli,~G.; Bouchaud,~J.~P.; Miyazaki,~K.; Reichman,~D.~R. {Inhomogeneous
  Mode-Coupling Theory and Growing Dynamic Length in Supercooled Liquids}.
  \emph{Phys. Rev. Lett.} \textbf{2006}, \emph{97}, 195701\relax
\mciteBstWouldAddEndPuncttrue
\mciteSetBstMidEndSepPunct{\mcitedefaultmidpunct}
{\mcitedefaultendpunct}{\mcitedefaultseppunct}\relax
\EndOfBibitem
\bibitem[Szamel and Flenner(2010)Szamel, and Flenner]{Szamel2010Diverging}
Szamel,~G.; Flenner,~E. {Diverging length scale of the inhomogeneous
  mode-coupling theory: A numerical investigation}. \emph{Phys. Rev. E}
  \textbf{2010}, \emph{81}, 031507\relax
\mciteBstWouldAddEndPuncttrue
\mciteSetBstMidEndSepPunct{\mcitedefaultmidpunct}
{\mcitedefaultendpunct}{\mcitedefaultseppunct}\relax
\EndOfBibitem
\bibitem[Curtis et~al.(2002)Curtis, Koss, and Grier]{Curtis2002Dynamic}
Curtis,~J.~E.; Koss,~B.~A.; Grier,~D.~G. {Dynamic holographic optical
  tweezers}. \emph{Optics Commun.} \textbf{2002}, \emph{207}, 169--175\relax
\mciteBstWouldAddEndPuncttrue
\mciteSetBstMidEndSepPunct{\mcitedefaultmidpunct}
{\mcitedefaultendpunct}{\mcitedefaultseppunct}\relax
\EndOfBibitem
\bibitem[Ciccotti et~al.(1979)Ciccotti, Jacucci, and
  McDonald]{Ciccotti1979Thought}
Ciccotti,~G.; Jacucci,~G.; McDonald,~I.~R. {"Thought-experiments" by molecular
  dynamics}. \emph{J. Stat. Phys.} \textbf{1979}, \emph{21}, 1--22\relax
\mciteBstWouldAddEndPuncttrue
\mciteSetBstMidEndSepPunct{\mcitedefaultmidpunct}
{\mcitedefaultendpunct}{\mcitedefaultseppunct}\relax
\EndOfBibitem
\bibitem[Bernu et~al.(1985)Bernu, Hiwatari, and Hansen]{Bernu1985Molecular}
Bernu,~B.; Hiwatari,~Y.; Hansen,~J.~P. {A molecular dynamics study of the glass
  transition in binary mixtures of soft spheres}. \emph{J. Phys. C}
  \textbf{1985}, \emph{18}, L371--L376\relax
\mciteBstWouldAddEndPuncttrue
\mciteSetBstMidEndSepPunct{\mcitedefaultmidpunct}
{\mcitedefaultendpunct}{\mcitedefaultseppunct}\relax
\EndOfBibitem
\bibitem[Bernu et~al.(1987)Bernu, Hansen, Hiwatari, and Pastore]{Bernu1987Soft}
Bernu,~B.; Hansen,~J.~P.; Hiwatari,~Y.; Pastore,~G. {Soft-sphere model for the
  glass transition in binary alloys: Pair structure and self-diffusion}.
  \emph{Phys. Rev. A} \textbf{1987}, \emph{36}, 4891\relax
\mciteBstWouldAddEndPuncttrue
\mciteSetBstMidEndSepPunct{\mcitedefaultmidpunct}
{\mcitedefaultendpunct}{\mcitedefaultseppunct}\relax
\EndOfBibitem
\bibitem[Donati et~al.(1998)Donati, Douglas, Kob, Plimpton, Poole, and
  Glotzer]{Donati1998Stringlike}
Donati,~C. et~al.  {Stringlike Cooperative Motion in a Supercooled Liquid}.
  \emph{Phys. Rev. Lett.} \textbf{1998}, \emph{80}, 2338--2341\relax
\mciteBstWouldAddEndPuncttrue
\mciteSetBstMidEndSepPunct{\mcitedefaultmidpunct}
{\mcitedefaultendpunct}{\mcitedefaultseppunct}\relax
\EndOfBibitem
\bibitem[Appignanesi et~al.(2006)Appignanesi, Rodr{\'\i}guez~Fris, Montani, and
  Kob]{Appignanesi2006Democratic}
Appignanesi,~G.~A.; Rodr{\'\i}guez~Fris,~J.~A.; Montani,~R.~A.; Kob,~W.
  {Democratic Particle Motion for Metabasin Transitions in Simple Glass
  Formers}. \emph{Phys. Rev. Lett.} \textbf{2006}, \emph{96}, 057801\relax
\mciteBstWouldAddEndPuncttrue
\mciteSetBstMidEndSepPunct{\mcitedefaultmidpunct}
{\mcitedefaultendpunct}{\mcitedefaultseppunct}\relax
\EndOfBibitem
\bibitem[Biroli and Bouchaud(2007)Biroli, and Bouchaud]{Biroli2007Critical}
Biroli,~G.; Bouchaud,~J.-P. {Critical fluctuations and breakdown of the
  Stokes--Einstein relation in the mode-coupling theory of glasses}. \emph{J.
  Phys.: Condens. Matter} \textbf{2007}, \emph{19}, 205101\relax
\mciteBstWouldAddEndPuncttrue
\mciteSetBstMidEndSepPunct{\mcitedefaultmidpunct}
{\mcitedefaultendpunct}{\mcitedefaultseppunct}\relax
\EndOfBibitem
\bibitem[Franz et~al.(2011)Franz, Parisi, Ricci-Tersenghi, and
  Rizzo]{Franz2011Field}
Franz,~S.; Parisi,~G.; Ricci-Tersenghi,~F.; Rizzo,~T. {Field theory of
  fluctuations in glasses}. \emph{Eur. Phys. J. E} \textbf{2011}, \emph{34},
  102\relax
\mciteBstWouldAddEndPuncttrue
\mciteSetBstMidEndSepPunct{\mcitedefaultmidpunct}
{\mcitedefaultendpunct}{\mcitedefaultseppunct}\relax
\EndOfBibitem
\bibitem[Franz et~al.(2010)Franz, Parisi, Ricci-Tersenghi, and
  Rizzo]{Franz2010Properties}
Franz,~S.; Parisi,~G.; Ricci-Tersenghi,~F.; Rizzo,~T. {Properties of the
  perturbative expansion around the mode-coupling dynamical transition in
  glasses}. \textbf{2010}, arXiv:1001.1746. arXiv.org e-Print
	archive. http://arxiv.org/abs/1001.1746 (accessed Jun 2010). \relax
\mciteBstWouldAddEndPunctfalse
\mciteSetBstMidEndSepPunct{\mcitedefaultmidpunct}
{}{\mcitedefaultseppunct}\relax
\EndOfBibitem
\end{mcitethebibliography}
\providecommand*\mcitethebibliography{\thebibliography}
\csname @ifundefined\endcsname{endmcitethebibliography}
  {\let\endmcitethebibliography\endthebibliography}{}

\clearpage

%
%
%


\end{document}